# Electron-Induced State Conversion in Diamond NV Centers Measured with Pump-Probe Cathodoluminescence Spectroscopy

*Magdalena Solà-Garcia[†], Sophie Meuret[†], Toon Coenen[†,§], and Albert Polman[†]*

[†]Center for Nanophotonics, AMOLF, Science Park 104, 1098 XG, Amsterdam, The Netherlands
[§]Delmic BV, Kanaalweg 4, 2628 EB Delft, The Netherlands

**Abstract**

Nitrogen-vacancy (NV) centers in diamond are reliable single-photon emitters, with applications in quantum technologies and metrology. Two charge states are known for NV centers: $NV^0$ and $NV^-$, with the latter being mostly studied due to its long electron spin coherence time. Therefore, control over the charge state of the NV centers is essential. However, an understanding of the dynamics between the different states still remains challenging. Here, conversion from $NV^-$ to $NV^0$ due to electron-induced carrier generation is shown. Ultrafast pump-probe cathodoluminescence spectroscopy is presented for the first time, with electron pulses as pump, and laser pulses as probe, to prepare and read out the NV states. The experimental data is explained with a model considering carrier dynamics (0.8 ns), $NV^0$ spontaneous emission (20 ns) and $NV^0 \rightarrow NV^-$ back transfer (500 ms). The results provide new insights into the $NV^- \rightarrow NV^0$ conversion dynamics, and into the use of pump-probe cathodoluminescence as a nanoscale NV characterization tool.

Nitrogen-vacancy (NV) centers in diamond are promising elements for quantum optical systems since they are single-photon emitters[1,2] with high photostability, quantum yield and brightness, even at room temperature.[3–6] Moreover, they are integrated inside a wide-bandgap solid-state host, the diamond lattice, making them robust against decoherence and allowing device scalability.[7–9] NV centers exhibit two different configurational states, the $NV^0$ state, with a zero-phonon line (ZPL) at 2.156 eV (λ = 575 nm), and the $NV^-$ state, with a ZPL at 1.945 eV (λ = 637 nm).[2] NV centers in the $NV^-$ state have received most of the attention in the past years since they exhibit a long electron spin coherence time that can be optically manipulated and read out,[9,10] which, together with the characteristics mentioned previously, make them suitable as building blocks for quantum technologies,[9,11,12] nanoscale magnetometry,[13,14] and other applications.[15,16] Typically, synthetically prepared diamonds with NV centers contain both $NV^0$ and $NV^-$ states. Previous work has shown that the state of an NV center can be converted from $NV^-$ to $NV^0$ (*ionization*) and vice versa (*recombination*). For example, the state of the NV centers can be changed by laser irradiation[17–19], as well as by shifting the Fermi level, either chemically,[20–22] or by applying an external voltage[23,24].

Overall, the control and understanding of NV state dynamics is key to the development of efficient quantum optical systems based on NV centers.

So far, most work on NV characterization and state conversion dynamics has focused on optical excitation and readout of the NV state. However, NV centers can also be excited by high-energy (1-200 keV) electrons, using either a scanning or transmission electron microscope (SEM or TEM), while the emitted cathodoluminescence (CL) is collected. Given the small electron beam spot size, the study of NV centers with electron excitation allows for a spatial resolution only limited by the diffusion of carriers, which can be down to the nanometer scale[25]. This opens the possibility to excite directly NV centers in nanodiamonds with high spatial resolution[26] and study the coupling of locally-excited nanostructures to NV centers[27,28], among others. Furthermore, NV centers are good platforms to study the fundamentals of quantum optics with electrons, in contrast to optical measurements. Electron-beam excitation of NV centers involves a multi-step process, in which the primary electron beam inelastically interacts with the diamond lattice, creating bulk plasmons that decay by generating charge carriers.[29–31] These carriers then diffuse through the diamond and recombine, partially through the excitation of NV centers. Single-photon emission of individual NV centers excited with electrons has already been demonstrated using measurements of the CL photon autocorrelation function ($g^{(2)}$).[26] Interestingly, in CL experiments typically only emission from the $NV^0$ state is observed,[25,26,32–37] with one exception[32] in which a very small $NV^-$ CL signal was observed at low temperature (16 K). This raises the question whether (1) the electron beam does not excite NV centers in the $NV^-$ state, (2) the electron beam quenches the $NV^-$ transition, or (3) the electron beam converts NV centers from the $NV^-$ to the $NV^0$ state. Answering this question is essential to understand the NV state dynamics in general, and to further exploit the use of CL in nanoscale characterization of atomic defects acting as single-photon emitters.

In this paper we study the interaction of electrons with NV centers, and in particular their state conversion dynamics. We perform the experiments using pump-probe CL spectroscopy, a novel technique that allows to study excited state dynamics at ultrafast timescales. Previous works combining electron and light excitations in a TEM include photon-induced near-field electron microscopy (PINEM)[38,39], in which the electron gains or loses energy when interacting with the optical-induced near-field, and femtosecond Lorentz microscopy[40], in which the laser-induced magnetization dynamics are probed with the electrons. Similarly, photo-induced carrier dynamics have been studied in an SEM by analyzing the secondary electron yield after laser excitation[41]. However, in these configurations the electron acts as a probe, since the signal is either transmitted or secondary electrons. In contrast, in pump-probe CL the final signal is the emitted light, either CL or photoluminescence (PL), therefore the electron can also act as a pump. In this work, we use an ultrafast SEM in which picosecond electron pulses are used to pump the diamond sample, while synchronously we optically probe the NV state. The electron pulses are generated using a laser-driven cathode configuration, a technique initially demonstrated by Merano et al. using a gold cathode[42], and further developed in combination with field-emission guns (FEGs) to improve the spatial and temporal resolution[43,44]. After ultrafast excitation of the NV centers, the CL and PL spectra are collected for spectral and temporal characterization. We find that repeated pulsed electron excitation (5.04 MHz) causes a state conversion from $NV^-$ to $NV^0$, until a steady state is achieved in which the electron-induced $NV^- \rightarrow NV^0$ conversion is balanced by the reverse $NV^0 \rightarrow NV^-$ back transfer. The steady state $NV^0$ population under electron irradiation can be controlled by the number of electrons per pulse. We describe the results with a model that includes electron-induced carrier generation and diffusion, with the NV centers acting as carrier traps and electrons converting NV centers from the $NV^-$ to the $NV^0$ state. The time dynamics of carrier diffusion (~0.8 ns), $NV^0$ decay (~ 20 ns) and $NV^0 \rightarrow NV^-$ back transfer (~ 500 ms) are clearly observed from the pump-probe transients.

**Pump-probe CL setup**

The pump-probe CL experiments are performed inside a SEM. We focus the 4$^{th}$ harmonic (λ=258 nm) of an Yb-doped fiber fs-laser on the electron gun to generate electron pulses by photoemission[42,45] (Figure 1a). Photoemission of electron pulses using this setup was characterized previously,[46] showing that the generated electron pulses are in the picosecond regime, similar to other work.[44,47] The electron beam is focused on a single spot on the sample, corresponding to the center of the area irradiated by the laser beam. We synchronously excite the sample at the electron-irradiated region with 2$^{nd}$ harmonic (λ=517 nm) pulses generated by the same fs laser, which are focused inside the SEM chamber to a ~10 μm-diameter spot on the sample using an Al parabolic mirror. The 2$^{nd}$ harmonic path length can be tuned within a ± 2 ns time window, such that the optical excitation pulse on the sample is delayed (or advanced) with respect to the electron pulse. CL and PL are collected by the parabolic mirror and directed to either a spectrometer or a time-correlated single photon counting (TCSPC) module. We use a 300 μm thick single-crystal diamond sample (obtained from Element 6 Inc.), grown by chemical-vapor deposition (<1 ppm nitrogen concentration, <0.05 ppm boron concentration), containing an approximate NV concentration of $[NV_{tot}]$=1.2 ppb (200 μm$^{-3}$). The sample is coated with a thin charge dissipation layer (E-spacer 300) to avoid charging when exciting with the electron pulses.

**CL, PL and pump-probe measurements**

Using the pump-probe CL setup, we acquire first PL and CL spectra, shown in Figure 1b. The PL spectrum shows emission from the ZPL of $NV^-$ (λ=637 nm) and $NV^0$ (λ=575 nm), with both ZPL transitions accompanied by phonon replicas, forming a broadband spectrum in the 575-800 nm spectral range. A Raman peak at λ=555 nm is also observed,[48] as well as a peak around 563 nm, which has been observed in previous work and preliminarily attributed to a divacancy defect.[37,49,50] The CL spectrum, obtained when exciting with a 5 keV pulsed electron beam clearly shows the ZPL of the $NV^0$ state, with phonon sideband, but no emission from the NV⁻ state is observed, similar to previous work.[25,26,32–37] The relative contribution of NV⁻ and NV⁰ states to the PL spectrum is obtained by a fitting procedure, with the CL spectrum as a reference for the spectral shape of the NV⁰ emission (see Supporting information). Using estimated optical absorption cross sections at the laser excitation wavelength (see Supporting information) we derive the NV⁻ and NV⁰ fractions: $[NV^-]/[NV_{tot}] \approx 0.4$ and $[NV^0]/[NV_{tot}] \approx 0.6$.

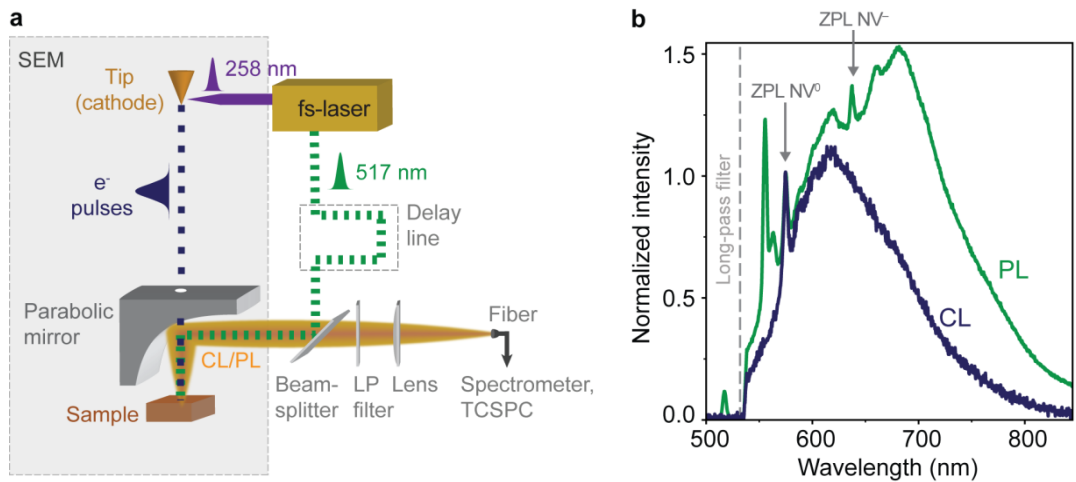

**Figure 1.** Pump-probe CL setup and NV centers spectra. **(a)** Schematic of the pump-probe CL setup. The 4th harmonic (λ=258 nm) of a fs laser is focused on the electron cathode to induce photoemission of electron pulses (0-400 electrons/pulse, picosecond temporal spread). The 2nd harmonic (λ=517 nm) of the same laser synchronously excites the sample to readout the NV state. The light pulse is delayed 1.3 ns with respect to the electron pulse. The emitted light, CL, PL or both, is collected using a parabolic mirror and analyzed with a spectrometer or TCSPC module. A long-pass (LP, λ>532 nm) filter is used to remove the light from the excitation laser. **(b)** Photoluminescence (green) and cathodoluminescence (blue) spectra obtained independently when exciting a bulk diamond sample with either 517 nm pulsed laser beam (0.9 nJ/pulse) or a 5 keV pulsed electron beam (400 electrons/pulse), respectively. Both spectra are obtained when exciting with a repetition rate of 5.04 MHz and at the same position on the sample. CL and PL spectra have been normalized by the amplitude of the NV⁰ ZPL at 575 nm.

Our pump-probe measurements consist of the independent acquisition of a set of spectra: only CL, only PL, and pump-probe (PP). The latter is obtained under simultaneous electron and light excitation, with the light pulse arriving 1.3 ns after each electron pulse. A set of spectra is shown in Figure 2a. All measurements were performed at the same spot on the sample, to avoid effects due to concentration inhomogeneities. In addition to the differences in the PL and CL spectra mentioned above, we also observe that the PL signal is an order of magnitude higher than the CL one. Even though a detailed comparison between both magnitudes is complex due to the different incident powers and excitation mechanisms, we can estimate the number of NVs excited in each case. The laser spot size has a diameter of around 10 μm and large penetration depth, due to the low absorption of diamond and low NV concentration. Therefore, the volume is mostly determined by the collection volume of the setup (see Methods). Instead, the primary interaction volume of the 5 keV electron beam is around 0.4 µm³, as calculated from Monte Carlo-based simulations using the software Casino[51]. Even though the effective volume is enlarged due to carrier diffusion, as will be shown below, it is still smaller than the volume excited by the laser. A sketch of both volumes is shown in Fig. 2c. Taking into account the optical cross-sections and collection geometry, we estimate that we collect PL from around $1.4 \times 10^4$ NVs per pulse for an incident power of 0.9 nJ (per pulse). Comparing the magnitude of the PL and CL signals, we can also extract that an average of 900 NV centers in the $NV^0$ state are excited per electron pulse, in the steady state situation, as will be discussed further on. In this case, each electron pulse contained 400 electron with 5 keV energy (corresponding to 0.32 pJ per pulse).

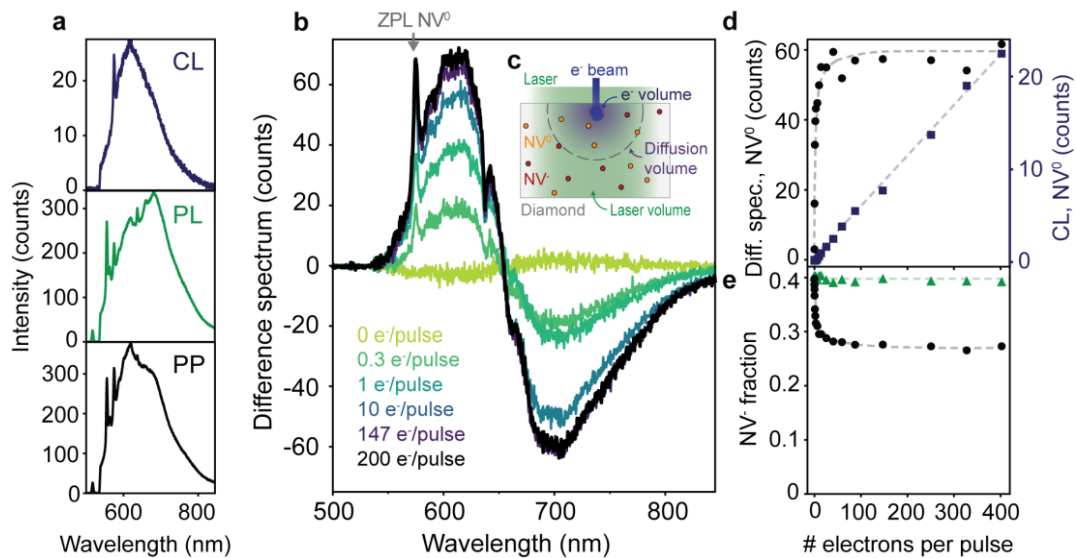

**Figure 2.** $NV^- \rightarrow NV^0$ conversion under electron excitation **(a)** Top: CL spectrum (5 keV, 400 electrons/pulse), middle: PL spectrum (λ=517 nm, 0.9 nJ/pulse), bottom: pump-probe (PP) spectrum obtained when both electrons and light (same conditions as before) excite the sample (5.04 MHz). The acquisition time was 1 min in all cases. **(b)** Difference spectrum, obtained by subtracting CL and PL spectra from the PP spectrum. **(c)** Sketch of the laser and electron excitation on the sample, representing the different volumes of primary electron interaction, diffusion of carriers and laser volume. **(d)** $NV^0$ ZPL intensity (λ=575 nm) of the difference spectrum (black circles) and from the CL-only spectrum (blue squares) as a function of the average number of electrons per pulse. The $NV^0$ ZPL of the difference spectrum shows saturation at around 20 electrons/pulse, while in the case of CL the dependence is linear. Dashed lines are shown as guides for the eye. **(e)** $NV^-$ fraction obtained from the PP as a function of the number of electrons per pulse. The green triangles indicate the $NV^-$ fraction derived from the PL spectra (all at the same PL pump power). Dashed lines are guides for the eye.

Using the PL, CL and PP spectra shown above, we can analyze the effect of electron irradiation on NV centers. We define the quantity of difference spectrum, obtained when subtracting CL and PL spectra from the PP spectrum. This analysis allows to study the correlation between electron and light excitation of the NV centers. Therefore, no correlation would lead to a flat difference spectrum. Instead, the difference spectrum obtained from the data in Figure 2a exhibits clear features, as shown in Figure 2b (black curve). We observe an increase of the signal (positive counts) in the lower-wavelength spectral band, corresponding to the $NV^0$ emission. As a reference, we observe a clear peak corresponding to the $NV^0$ ZPL. We also observe a concomitant decrease in the longer-wavelength band, corresponding to $NV^-$ emission. In this case, the $NV^-$ ZPL is visible as a dip. This implies that after electron excitation the number of emitting $NV^0$ centers is increased, while the number of $NV^-$ centers is decreased. The results suggest that centers in the $NV^-$ states are converted into $NV^0$ states under electron irradiation, corresponding to hypothesis (3) exposed earlier in the text. Difference spectra derived for different sets of measurements at 0.3, 1, 10 and 147 electrons per pulse are also shown in Figure 2b, as well as a reference measurement (no electron irradiation). Each set of measurements corresponds to the acquisition of independent CL, PL and PP spectra, in which the number of electrons per pulse is varied, while keeping the laser excitation power constant at 0.9 nJ per pulse. We again observe $NV^- \rightarrow NV^0$ conversion, with the number of converted centers rising for increasing average number of electrons per pulse. This behavior in the difference spectra was consistently observed in other measurements at different areas of the sample, and also with other electron energies (30 keV, Fig. S1).

To further investigate the electron-induced $NV^- \rightarrow NV^0$ conversion trend, we plot the amplitude of the $NV^0$ ZPL as a function of the number of electrons per pulse (Figure 2d). Saturation of the signal from the $NV^0$ ZPL is observed above ~20 electrons per pulse, suggesting that this is the required electron flux (at 5.04 MHz) to induce the saturation of the $NV^-$ conversion in the volume of the sample excited by electrons. For reference, Figure 2d also shows the CL intensity for the $NV^0$ ZPL as a function of the number of electrons per pulse. The plot shows a linear trend, indicating that the $NV^0$ CL signal is not saturating with increasing electron dose, i.e., there is no strong depletion of the ground state population. Therefore, from these results we derive that electrons can either excite NV centers in the $NV^0$ state, which leads to a linear dependence on the electron flux, or convert $NV^-$ into $NV^0$, which saturates with increasing number of electrons per pulse.

From the data in Figure 2b we can also derive the $NV^-$ population as a function of the number of electrons per pulse, as plotted in Figure 2e. This derivation is done by fitting the $NV^0$ and $NV^-$ contributions from the PP measurements (see Supporting information). Starting from the initial $NV^-$ fraction of 0.4 for the reference measurement, as already derived before, the population of centers in the $NV^-$ state rapidly decreases with increasing number of electrons per pulse, reaching a saturation level corresponding to 0.26 $NV^-$ fraction. We attribute this saturation level to the full conversion of $NV^-$ centers into $NV^0$ centers within the volume excited by the electrons, as will be discussed further on. The fact that the $NV^-$ fraction does not reach zero at saturation is attributed to the difference between excitation and collection volumes of electron and laser beam, as sketched in Figure 2c. For completeness, in Figure 2e we also show the $NV^-$ fraction derived from the PL measurements taken in each set of measurements from Figure 2b. We observe that the $NV^-$ fraction under only laser irradiation remains approximately constant, meaning that the $NV^-$ population before each set of measurements is identical. The fact that the $NV^-$ population is unchanged also implies that the electron-induced $NV^-{\rightarrow}NV^0$ conversion is reversible, i.e., there is an $NV^0{\rightarrow}NV^-$ back transfer process, and that damage induced by the electron to the sample is negligible. Given that $NV^- \leftrightarrow NV^0$ conversion has also been observed due only to laser irradiation[17–19], we also acquired PL spectra at different incident powers. The results are presented in Figure S2 and show that the $NV^-$ fraction remains constant for increasing laser power, therefore proving that NV conversion due to only laser irradiation is negligible in our experiment. Pump-probe measurements with different delays between electron and light were also acquired (Figure S3), but no significant differences are observed. This is attributed to the fact that the $NV^0{\rightarrow}NV^-$ back transfer is on the order of milliseconds, as will be demonstrated below, larger than the time between pulses (198 ns at 5.04 MHz).

**Excitation, emission and conversion dynamics**

In order to further describe the interaction of electrons with NV centers, we study the excitation and emission dynamics of NV centers at the nanosecond timescale, as well as the $NV^0{\rightarrow}NV^-$ back transfer that occurs in the millisecond scale. The time-dependent CL emission from NV centers upon electron excitation is shown in Figure 3a, which has been measured using the TCSPC technique. Notice that the CL intensity corresponds only to emission from excited $NV^0$ centers, given that $NV^-$ emission is not probed with CL. The CL signal exhibits a gradual increase in the first 2 ns, reaching a maximum emission at around 2.2 ns (see inset). We ascribe this initial increase to the diffusion of carriers beyond the primary electron-excited volume, which increases the excited $NV^0$ population well after the initial ps-electron pulse excitation. After the first 2 ns we observe a decay of the CL intensity, from which we extract a characteristic decay time of ~20 ns, in agreement with the typical radiative decay time of excited $NV^0$ centers.[26,52] We also observe a ~100 ps spike at 0 ns, which accounts for around 1% of the total intensity. The origin of this fast decay is unknown. The intensity of this peak depends on the position on the sample, as well as electron energy. Nevertheless, the amplitude of this peak does not show any correlation with the magnitude of the $NV^-{\rightarrow}NV^0$ conversion, from which we infer that both effects are unrelated.

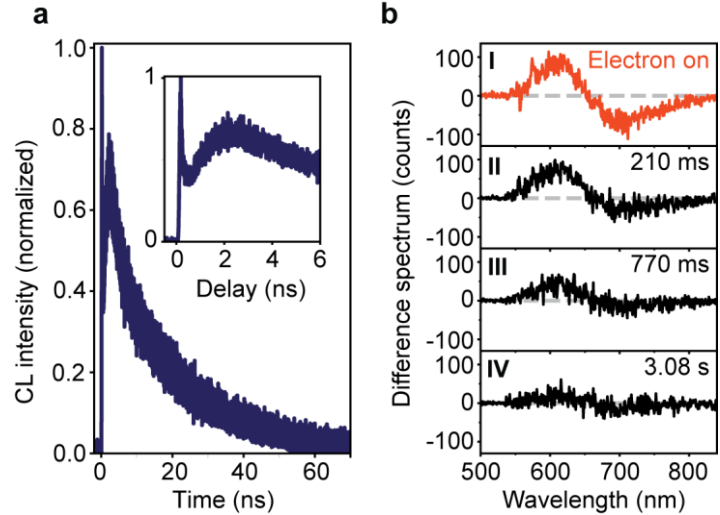

**Figure 3.** Carrier diffusion, excitation and back transfer dynamics. **(a)** Peak-normalized CL intensity upon pulsed electron excitation (5 keV, ~450 electrons/pulse, 5.04 MHz) at t=0 ns, measured with time-correlated single photon counting. Data are taken in the $NV^0$ 575-725 nm spectral band. Inset: enlarged early-timescale. **(b)** Difference spectrum (defined as PP–CL–PL) obtained with the electron beam on (I) and 210 ms, 770 ms and 3.08 s after the electron beam was blanked (II-IV, respectively). The $NV^0 \rightarrow NV^-$ back transfer takes around 500 ms. The time-resolution of this experiment is 70 ms.

In contrast to the fast carrier diffusion and $NV^0$ emission dynamics, previous studies of optically induced $NV^0 \leftrightarrow NV^-$ conversion suggest that the $NV^0 \rightarrow NV^-$ back transfer is in the millisecond regime[17]. To study this, we performed time-resolved spectral measurements over a millisecond time scale. We used the minimum exposure time possible in our spectrometer, acquiring a spectrum every 70 ms. The repetition rate is kept at 5.04 MHz, as in the previous experiments. We performed a spectral acquisition sequence in which initially both the electron and laser beam were irradiating the sample (PP spectrum). At some point during the acquisition, the electron beam was blanked, while the laser continued exciting the sample, and spectra kept being collected every 70 ms. In this way, the NV population can be probed immediately after the electron beam is switched off. Afterwards, we also acquired CL and PL spectra with the same exposure time, such that a difference spectrum can be derived, similar to Figure 2b. An example of the obtained difference spectrum is shown in Figure 3b-I, which again reflects the $NV^- \rightarrow NV^0$ conversion by the electron-excited carriers. In this case, the electron beam was still irradiating the sample. Figure 3b-II shows the difference spectrum obtained 210 ms after switching off the electron beam. Notice that here the difference spectrum is obtained by subtracting only PL from the PP measurement, given that there is no CL. We observe a 30% decrease of the intensity of the difference spectrum, indicating that most of the converted $NV^-$ centers still remain in the $NV^0$ state, and only some have converted back into $NV^-$. Results after 770 ms and 3.08 s are also plotted (Figure 3b-III,IV), in which we observe a progressive decay of the signal, indicating that $NV^0$ centers are converted back to the $NV^-$ state. A complete transient of the average signal in the difference spectrum as a function of time is provided in Fig. S4. These data indicate that the electron-induced $NV^- \rightarrow NV^0$ state conversion is reversible, with the back transfer taking place within a characteristic time of ~500 ms. This time scale is in agreement with earlier work, in which back transfer of optically-induced NV⁻→NV⁰ conversion, was found to occur with a characteristic time of 465 ms.[17]

**Discussion and phenomenological model**

Optically-induced state conversion from NV⁻ to NV⁰ has been previously explained to take place by the release of an electron from the NV⁻ center to the conduction band of diamond.[19,52–54] Literature values for the difference in energy between the NV⁻ ground state and the conduction band range from 2.6 to 4.3 eV,[17,19,52] and the NV⁻/NV⁰ optical conversion typically requires a two-photon absorption process. In our experiment, we propose a model in which electron-hole pairs generated from the electron cascade can recombine, thus providing the energy to induce the release of the bound electron from the NV⁻ center, given that the bandgap of diamond is 5.5 eV. This conversion mechanism is similar to that in optical experiments, with the difference that the energy is provided by a carrier recombination event instead of two pump photons. This model is in agreement with previous work in which emission only from the NV⁰ state was observed when exciting with far-UV photons (λ=170 nm, above the bandgap of diamond)[55] and in electroluminescence[56,57]. In both cases, charge carriers are generated and NV centers are excited through the recombination of carriers, similar to CL. In addition to this, the energy provided by a single carrier recombination event is larger than the energy needed to induce the NV⁻→NV⁰ conversion, suggesting that a single carrier recombination event could already release the electron, without the need to first excite the NV⁻ center as in the case of optical experiments.[19,52–54] The latter suggestion requires further studies in the mechanism of NV⁻→NV⁰ conversion by carrier recombination, which are beyond the scope of this paper.

To qualitatively analyze the data shown above we model the electron-induced NV⁻→NV⁰ state conversion by means of a three-dimensional model, considering carrier diffusion and NV center conversion and excitation. We start by modelling the dynamics in the nanosecond regime, corresponding to carrier diffusion and NV⁰ decay. We use Monte Carlo simulations, using the software Casino[51], to obtain the three-dimensional spatial distribution of inelastic scattering events of the primary 5 keV electron beam. Most of the energy lost by the electron corresponds to the generation of bulk plasmons, described as excitations of the outer shell electrons[29], with an energy corresponding to 31 eV for diamond[30]. We then model the initial carrier distribution with a 3D Gaussian distribution, with standard deviation σ=0.185 μm estimated from the plasmon distribution derived from Casino simulations, and amplitude proportional to the number of electrons per pulse. We assume that each bulk plasmon effectively generates an average of 2 electron-hole pairs.[30] The concentration of charge carriers as a function of time and space ($\rho_{eh}(r,t)$) is then obtained by solving the diffusion equation, with carrier recombination described with a lifetime $\tau_R$.

Taking into account carrier diffusion, we model the concentration of NV⁻ in the ground state ($\rho_-$) and NV⁰ in the ground ($\rho_0^g$) and excited ($\rho_0^e$) states by means of a rate equation model:

$$\frac{\partial \rho_-(r,t)}{\partial t} = -v_{th}\, \rho_{eh}(r,t)\, \sigma_c^{eh}\, \rho_-(r,t) + \frac{\rho_{-i} - \rho_-(r,t)}{\tau_{back}} \tag{1a}$$

$$\frac{\partial \rho_0^g(r,t)}{\partial t} = v_{th}\, \rho_{eh}(r,t) \left[\sigma_c^{eh}\rho_-(r,t) - \sigma_0^{eh}\rho_0^g(r,t)\right] + \frac{\rho_0^e(r,t)}{\tau_0} - \frac{\rho_{-i} - \rho_-(r,t)}{\tau_{back}} \tag{1b}$$

$$\frac{\partial \rho_0^e(r,t)}{\partial t} = v_{th}\, \rho_{eh}(r,t)\, \sigma_0^{eh}\rho_0^g(r,t) - \frac{\rho_0^e(r,t)}{\tau_0} \tag{1c}$$

where $v_{th}$ is the thermal velocity of carriers, $\sigma_0^{eh}$ is the cross section to excite NV⁰ states by carriers, $\sigma_c^{eh}$ is the NV⁻→NV⁰ conversion cross section, $\tau_0$ is the lifetime of excited NV⁰ state, $\tau_{back}$ accounts for the NV⁰→NV⁻ back transfer, and $\rho_{-i}$ is the initial uniform concentration of NV⁻. In this model we assume that NV⁰ states can be excited by carriers, but NV⁻ states cannot, given that we do not observe NV⁻ signal in the CL measurements. Moreover, the interaction of the primary electron beam (picosecond temporal spread) with the sample, including generation of bulk plasmons and decay into carriers, is treated as instantaneous, given that it is much shorter than the characteristic time scale of the dynamics in Equations 1a-c.

Numerically solving the system of differential equations over time, and integrating $\rho_0^e(r,t)$ over the collection volume, allows to fit the trend in the first 2 ns of the time-dependent CL intensity shown in Figure 3a. The carrier lifetime derived from the fit is $\tau_R = 0.8\, ns$, corresponding to a diffusion length of 0.9 μm, which is in agreement with values reported for samples with a similar concentration of NV centers.[58] From the model we also find that excitation with 400 electrons (5 keV) leads to about 740 NV⁰ centers excited per pulse, close to the value independently derived from the comparison of PL and CL intensities in Figure 2a,b. Taking into account the obtained carrier lifetime, in Figure 4a we plot the spatial distribution of the carrier concentration at t=0 ns (solid black) and after 1 and 5 ns (dashed dark green and dotted light green, respectively), obtained from the expression of $\rho_{eh}(r,t)$ (Equation S1). The carrier distribution rapidly spreads out due to diffusion, with the total amount of carriers decreasing as a result of carrier recombination.

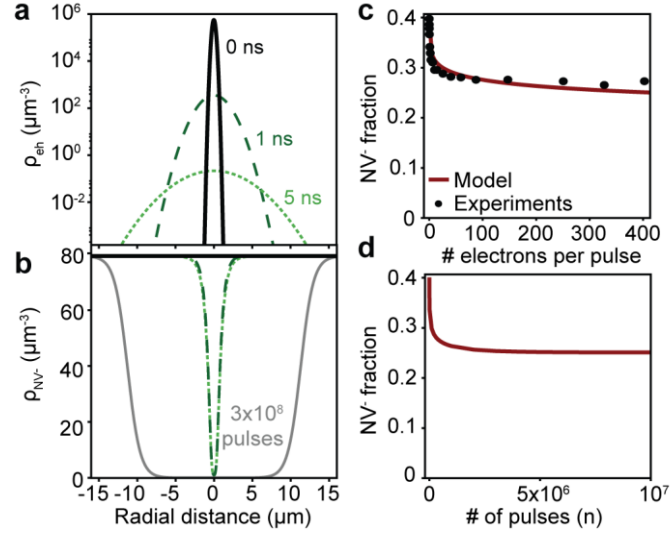

**Figure 4.** Carrier diffusion and rate equation models. **(a)** Spatial distribution of the concentration of carriers at t=0, 1 and 5 ns (solid black, dashed dark green and dotted light green, respectively). **(b)** Initial spatial distribution of the concentration of NV⁻ states (solid black) and after 1 and 5 ns (dashed dark and dotted light green) after a single electron pulse. The spatial distribution of NV⁻ states after 3×10⁸ pulses, corresponding to a typical acquisition time (~1 min), is also plotted (solid gray). **(c)** Modelled NV⁻ fraction as a function of the number of electrons per pulse (dark red curve), together with the experimental data (black circles). **(d)** NV⁻ fraction as a function of the number of pulses (400 electrons/pulse), obtained using the discrete rate equation model.

The calculated spatial distribution of the NV⁻ concentration is shown in Figure 4b, again at t=0, 1 and 5 ns, obtained by solving Equations 1a-c. Given that the electron excitation cross sections for NV⁰ excitation and NV⁻→NV⁰ conversion are unknown, we estimate them by considering the known exciton capture cross section of a nitrogen impurity in diamond,[59] $\sigma_0^{eh} = \sigma_c^{eh} = 3 \times 10^{-6}\ \mu m^2$. We consider $v_{th} = 100\ \mu m/ns$, $\tau_{back} = 500\ ms$, as obtained from the experimental data in Figure 3b, and an initial homogeneous NV⁻ fraction of 0.4 (black line in Figure 4b for t=0 ns), corresponding to the experimental data in Figure 2e. We observe that 1 ns after the first pulse, NV centers in the NV⁻ state that are located within a 1 μm range from the initial electron cascade have been converted to NV⁰ due to the interaction with carriers. For larger times (5 ns) the distribution of converted NV⁻ centers is nearly the same as for t=1 ns, as nearly all carriers have recombined.

In order to account for longer time scales, corresponding to the back transfer from NV⁰ to NV⁻ and the time of acquisition of our experiments (typically 1 min, ~3×10⁸ pulses), we developed a discrete rate equation model. In this case, the concentration of NV⁻ centers is modelled as a function of the pulse number (n):

$$\rho_-(r,n) = \rho_-(r,0)\frac{\beta+\alpha(r)[1-\alpha(r)-\beta]^n}{\alpha(r)+\beta} \qquad (2)$$

where

$$\alpha(r) = 1 - e^{-v_{th}\sigma_c \int_0^T \rho_{eh}(r,t)\, dt} \qquad (3)$$

is the probability of carrier-induced conversion of centers in the NV⁻ states between subsequent pulses, with $T$ being the time between pulses (198 ns at 5.04 MHz), and

$$\beta = 1 - e^{-T/\tau_{back}} \qquad (4)$$

is the probability that an NV⁰ center transfers back to the NV⁻ state, again between subsequent pulses (see Supporting information). Using this model, in Figure 4b we plot the spatial distribution of NV⁻ centers after $3\times10^8$ pulses (solid gray), corresponding to a typical acquisition time (1 min), in which steady state has been reached. The calculated steady state NV⁻ fraction as a function of the number of electrons per pulse is shown in Figure 4c, which is overlaid with the experimental data from Figure 2e (black circles). Each point in the plot corresponds to the steady state value calculated using Equation 2, and integrated over the excitation and collection volume (see Supporting information). In our model, taking the parameters discussed above, the only fit parameter is the collection depth of the CL system, which is 23 µm for the best fit. This is a reasonable value given the confocal geometry of the CL/PL collection system (see Methods). Figure 4d shows the calculated NV⁻ fraction as a function of the number of pulses. We observe that the NV⁻ fraction saturates for $\sim 5\times10^6$ pulses (1 s), consistent with the fact that steady state is reached for a time longer than the NV⁰→NV⁻ back transfer time. Overall, the model qualitatively describes properly the experimental data, therefore giving further proof for the proposed electron-induced mechanism for NV⁻→NV⁰ conversion dynamics.

**Conclusions**

In conclusion, we have used pump-probe CL spectroscopy to show that high energy (5 keV) electron irradiation of NV centers induces a state conversion from the NV⁻ to the NV⁰ state. We show that the NV⁻ population decreases when increasing the number of electrons per pulse that excite the sample, until saturation is reached, which is attributed to the full conversion of the NV⁻ centers in the volume excited through the electrons. Experiments also show that the NV⁻→NV⁰ conversion is reversible, with a typical back transfer time of 500 ms. We present a three-dimensional rate equation model, considering diffusion of electron-generated charge carriers and taking into account the integrated effect of subsequent pulses, which qualitatively describes the experimental results. This work shows that NV⁻ centers are effectively converted to NV⁰ centers by electron irradiation, and explains why NV⁻ emission is not observed in CL measurements. We envision that the pump-probe CL approach presented in this work can be applied to other complex solid-state emitter systems, to obtain further insight in their complex dynamical behavior.

**Methods**
**Ultrafast SEM.** A schematic of the setup is shown in Figure 1a. The pump-probe CL experiments are performed inside a SEM (Thermo Fisher Scientific/XL30 FEI) containing a Schottky field-emission electron cathode consisting of a ZrO coated W tip. The conditions used to generate the electron pulses are discussed in Ref.[46]. We use a diode-pumped Yb-doped fiber system (IMPULSE Clark-MXR) providing 250-fs light pulses at a wavelength of

λ=1035 nm and repetition rate of 5.04 MHz. The primary laser beam is guided through a harmonic generator to create 2nd, 3rd and 4th harmonics (517, 345 and 258 nm, respectively). The 4th harmonic is guided to the electron column and focused with a f=15 cm lens onto the electron cathode, which is accessible through a vacuum window. Earlier work using the same setup has shown that this photoemission process results in electron pulses with a temporal spread in the picosecond range.[46] We use a gradient neutral-density filter to change the 4th-harmonic pulse energy from 0 to 1.5 nJ/pulse, which results in an average number of electrons per pulse up to 400. The corresponding time-averaged beam current on the sample was 0-325 pA measured with a Faraday cup. The error in the current measurement is ~25%, limited by the stability in the laser power, and measurement method. In the experiments, the electron spot size has a diameter of ~600 nm. Using the same setup, a higher spatial resolution can be achieved at the expense of lower current on the sample.[46] All the experiments are performed at room temperature and at a pressure of $10^{-6}$ mbar.

**Laser-electron beam overlap.** The 2nd harmonic (λ=517 nm) of the same primary laser beam is passed through a linear stage (Newport M-IMS600PP) with motor controller (Newport ESP301-1G), after which it is sent through a pellicle beam splitter (8:92), guided into the SEM sample chamber through a vacuum window, and focused onto the sample to a ~10 μm-diameter spot using an Al parabolic mirror (1.46π sr acceptance angle, 0.1 parabola parameter and 0.5 mm focal distance). In the pump-probe measurements the 2nd harmonic path length was tuned such that the light pulse was delayed 1.3 ns with respect to the electron pulse. The 2nd and 4th harmonic laser powers were independently controlled such that measurements with varying number of electrons per pulse could be done for constant 2nd harmonic PL power.

**CL and PL collection.** Luminescence from the sample is collected using the Al parabolic mirror and directed to a light collection and analysis system. Light collected by the mirror is focused (f=16 cm) onto the entrance facet of a multimode fiber (550 μm core diameter) creating a confocal collection geometry, which limits the PL and CL collection depth in the sample. The fiber guides the light to a Czerny-Turner spectrometer equipped with a CCD array detector (Princeton Spec10) and grating containing 150 lines/mm and blaze wavelength corresponding to 500 nm. A long-pass filter (λ>532 nm) is used to suppress scattered pump laser light in the detection path. TCSPC measurements are performed by sending the CL signal to a single photon avalanche photodiode (MPD PD-100) analyzed by time correlation (Picoquant PicoHarp 300), which builds a delay histogram. In this case, an additional bandpass filter (λ=650±75 nm) is used, corresponding to the spectral range within which NV emission occurs. We use the 3rd harmonic laser pulse measured with a photodiode as the trigger for the time-correlated measurements. The PL, CL and PP data in Figure 1b and Figure 2a are collected over a time of 1 min each. The light collection geometry in this setup typically allows the collection of light within a 20x20 μm² area. Only light emitted in this area, and within the escape cone of diamond, can be collected efficiently. Given the critical angle for diamond ($\theta_c$<24.6°), we can estimate that light emitted at a depth down to 20 µm inside the diamond can still be collected. Nevertheless, emission beyond this 20 µm depth might reach the surface at a position outside of the collection area, thus the collection efficiency decreases at larger depths.

**Associated content**
*Supporting information. Additional experimental data and description of the model.*

**Acknowledgements**
This work is part of the research program of AMOLF which is partly financed by the Dutch Research Council (NWO). This project has received funding from the European Research Council (ERC) under the European Union's Horizon 2020 research and innovation programme (grant agreement No. 695343). We gratefully acknowledge the technical support of Erik Kieft, Hans Zeijlemaker and Dion Ursem. We also gratefully acknowledge discussions with Mayeul Chipaux in the initial phase of this project, Patrick Maletinksy and Elke Neu for providing the bulk diamond sample, and Kévin Cognée for careful reading of the manuscript.

# Supporting information

## Electron-Induced State Conversion in Diamond NV Centers Measured with Pump-Probe Cathodoluminescence Spectroscopy

*Magdalena Solà-Garcia[†], Sophie Meuret[†], Toon Coenen[†,§], and Albert Polman[†]*

[†]Center for Nanophotonics, AMOLF, Science Park 104, 1098 XG, Amsterdam, The Netherlands
[§]Delmic BV, Kanaalweg 4, 2628 EB Delft, The Netherlands

### 1. Pump-probe measurements: set of spectra at 30 keV

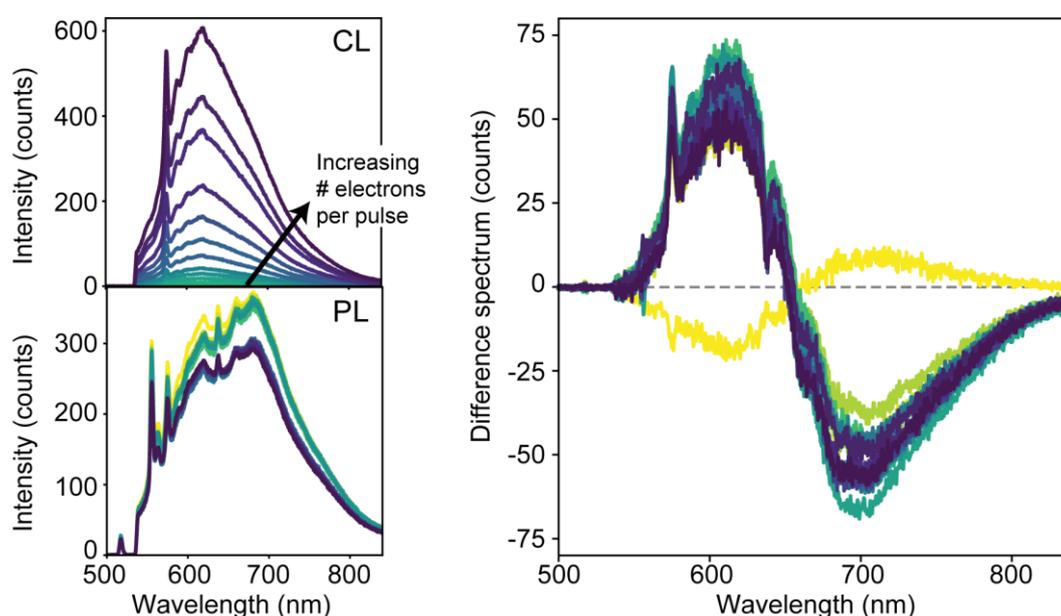

**Figure S1.** Pump-probe experiments performed with an electron energy of 30 keV. (a) Cathodoluminescence (CL), (b) Photoluminescence (PL) and (c) difference spectrum (PP − PL − CL, where PP stands for pump-probe). The colors of the curves indicate the number of electrons per pulse, going from 0 (yellow) up to 208 (dark purple). The difference spectrum reflects again the $NV^-$ →$NV^0$ conversion due to electron irradiation. Nevertheless, when performing these measurements at 30 keV we consistently observe deterioration of the sample after each CL measurement, as can be observed from the PL measurements (a) taken before each pump-probe measurement (and after each CL measurement). This deterioration of the sample can also be observed in the reference measurement (yellow curve, 0 electrons per pulse), which does not show a completely flat spectrum.

## 2. Laser power dependence of NV⁻ fraction

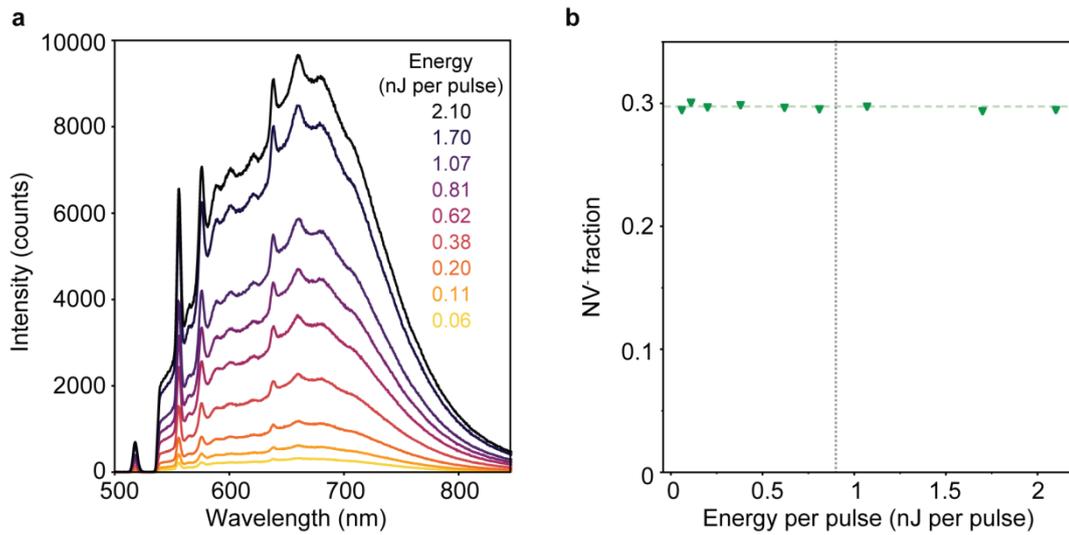

**Figure S2.** NV⁻ fraction as a function of the incident laser power. (a) PL spectra obtained at the same spot on the sample but different incident laser power, ranging from 0.06 up to 2.1 nJ per pulse. (b) NV⁻ fraction as a function of the energy per pulse, derived from the PL spectra in (a). The NV⁻ fraction remains constant for the different values of the incident power, indicating that optically-induced NV⁻→NV⁰ conversion (or vice versa) is negligible in this case. The dotted gray line indicates the power at which the experiments from Fig. 2 were performed, while the dashed green line serves as a guide for the eye. These measurements were all acquired at the same spot on the sample, but different from the spot in which measurements from Figure 2 were performed, thus explaining why the NV⁻ fraction is different in both cases.

## 3. Pump-probe measurements vs. electron-laser delay

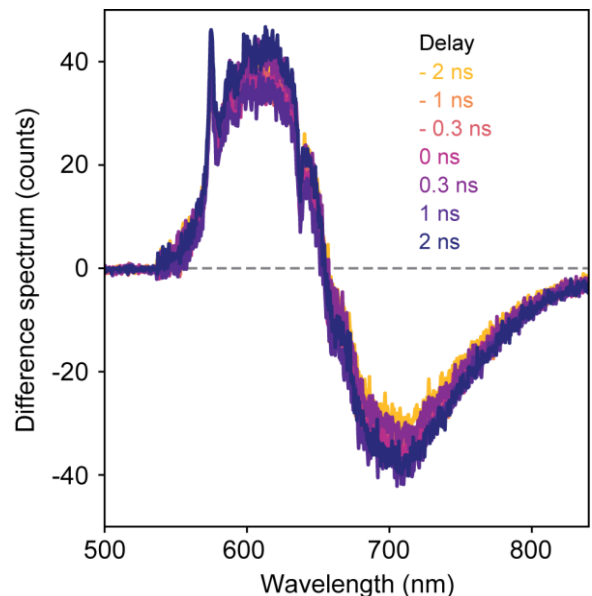

**Figure S3.** Difference spectrum obtained for different laser arrival times. The delay indicates the difference in arrival time of electrons and laser pulses to the sample, with negative delay indicating that the laser arrives before the electron beam. The NV⁻→NV⁰ conversion is again observed in the difference spectrum, but there are no differences among the different delays, due to the fact that the electron-induced conversion has a timescale in the millisecond regime (Fig. 3b), much larger than the time between pulses in the experiments (198 ns at 5.04 MHz).

## 4. Transient of the back transfer dynamics

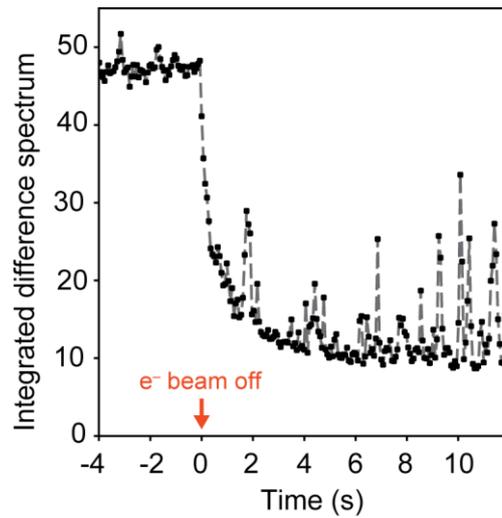

**Figure S4.** Temporal evolution of the $NV^0 \rightarrow NV^-$ back transfer after turning off the electron beam. The y-axis corresponds to the mean value of the intensity of each difference spectrum, after taking the absolute value. Each spectra is extracted every 70 ms during the measurement (Fig. 3b).

## 5. Data analysis

Absorption cross sections of $NV^0$ and $NV^-$ at the excitation wavelength of $\lambda=517$ nm are estimated to be $2\times10^{-17}$ cm$^2$ and $1.4\times10^{-17}$ cm$^2$, respectively. In order to estimate these cross sections, first the $NV^-$ and $NV^0$ contributions to the PL spectrum are disentangled by taking the normalized CL spectrum as the $NV^0$ spectral shape, and assuming that the remaining PL spectrum corresponds to the $NV^-$ contribution. We then consider complementarity between emission and absorption spectra, and normalize by known absorption cross sections at the ZPL of each center.[1] The amount of excited NVs in PL from Figure 2b is calculated by considering the NV concentration (200 µm$^{-3}$), and a light collection depth of 23 µm. The $NV^-$ population in the pump-probe measurements (Figure 2d) is obtained by fitting the PP spectra, after subtraction of the CL spectra, and considering the estimated absorption cross sections for the $NV^-$ and $NV^0$ states. The fitting of the NV spectra are performed with a total of 14 Gaussian functions: for each NV state ($NV^0$ or $NV^-$), one Gaussian function is used to fit the ZPL and 6 broader Gaussian functions are used to account for the phonon replica. We estimate a relative error in the calculation of the $NV^-$ population of <25%, due to uncertainties in the fitting procedure.

## 6. Model

*Electron cascade simulations.* The spatial distribution of the creation of bulk plasmons by the primary electron beam was obtained with the Monte Carlo-based simulation software Casino. We used a diamond density of 3.51 g cm$^{-1}$ and bulk plasmon energy of 31 eV.[2] The beam diameter is set to 600 nm. From the simulations we derive an average of 70 bulk plasmons created per electron.

*Model for diffusion of charge carriers.* The evolution in space and time of the concentration of electron-hole pairs, $\rho_{eh}(r,t)$, is obtained by solving the three-dimensional diffusion equation in spherical coordinates, which gives:

$$\rho_{eh}(r,t) = \frac{a\sigma^3}{(2Dt+\sigma^2)^{3/2}} e^{-\frac{t}{\tau_R}} e^{-\frac{r^2}{4Dt+2\sigma^2}} \tag{S1}$$

where $D$ is the carrier diffusion coefficient and $\tau_R$ the carrier lifetime, which accounts for the recombination of carriers. We consider $D = 1\ \mu m^2\ ns^{-1}$, as obtained from literature.[3,4] The parameters $a$ and $\sigma$ correspond to the amplitude and standard deviation of the 3D initial Gaussian distribution of carriers, derived from Casino simulations. In our case, $\sigma = 0.185$ μm and $a = 1404\ n_{el}$, where $n_{el}$ is the number of electrons per pulse. We do not consider the effect of the diamond surface on the diffusion equation and recombination of carriers.

*Discrete rate equation model.* The concentration of NV⁻ as a function of position, r, and number of pulse, n, described by Equation 2 is derived by solving the rate equation:

$$\rho_-(r, n+1) = \rho_-(r,n) - \alpha(r)\rho_-(r,n) + \beta[\rho_-(r,0) - \rho_-(r,n)] \tag{S2}$$

Here, $\alpha(r)$ is described with Equation 3 and is obtained by considering the change in $\rho_-(r,t)$ between subsequent pulses only due to NV⁻→NV⁰ conversion, i.e. $\alpha(r) = \rho_{-,conv}(r,T) - \rho_{-,conv}(r,0)$. This process is described with the rate equation:

$$\frac{\partial \rho_{-,conv}(r,t)}{\partial t} = -v_{th}\ \sigma_c\ \rho_{eh}(r,t)\ \rho_{-,conv}(r,t) \tag{S3}$$

Similarly, we derive the expression of $\beta = \rho_{-,back}(r,T) - \rho_{-,back}(r,0)$ from Equation 4 by considering the change in $\rho_-(r,t)$ during the time between two pulses only due to the of NV⁰→NV⁻ back transfer, which is obtained by solving the rate equation:

$$\frac{\partial \rho_{-,back}(r,t)}{\partial t} = -\frac{1}{\tau_{back}}[\rho_{-,back}(r,t) - \rho_{-i}] \tag{S4}$$

In this description we assume that the change in $\rho_-(r,t)$ due to the NV⁻→NV⁰ conversion and due to the NV⁰→NV⁻ back transfer are independent between subsequent pulses. This assumption is valid since the back transfer time (500 ms) is much longer than the time between pulses (198 ns). In order to compare the model with the experimental data, we calculate the steady state value, $\rho_-(r,\infty)$, and integrate over a cylindrical volume, with cross section corresponding to the Gaussian profile of the excitation beam ($\sigma_{laser}$=5 μm). Given that diamond is transparent at the excitation wavelength (λ=517 nm), absorption is only due to excitation of NV centers, thus NVs will be excited through the entire sample. Nevertheless, PL will only be effectively collected up to a certain depth, which becomes the fit parameter.